# Channel State Feedback Schemes for Multiuser MIMO-OFDM Downlink

Hooman Shirani-Mehr and Giuseppe Caire

### Abstract

Channel state feedback schemes for the MIMO broadcast downlink have been widely studied in the frequency-flat case. This work focuses on the more relevant frequency selective case, where some important new aspects emerge. We consider a MIMO-OFDM broadcast channel and compare achievable ergodic rates under three channel state feedback schemes: analog feedback, direction quantized feedback and "time-domain" channel quantized feedback. The first two schemes are direct extensions of previously proposed schemes. The third scheme is novel, and it is directly inspired by rate-distortion theory of Gaussian correlated sources. For each scheme we derive the conditions under which the system achieves full multiplexing gain. The key difference with respect to the widely treated frequency-flat case is that in MIMO-OFDM the frequency-domain channel transfer function is a Gaussian correlated source. The new time-domain quantization scheme takes advantage of the channel frequency correlation structure and outperforms the other schemes. Furthermore, it is by far simpler to implement than complicated spherical vector quantization. In particular, we observe that no structured codebook design and vector quantization is actually needed for efficient channel state information feedback.

H. Shirani-Mehr and G. Caire are with the Ming Hsieh Department of Electrical Engineering, University of Southern California, Los Angeles, CA 90089 USA. E-mail: shiranim@usc.edu, caire@usc.edu.

# I. INTRODUCTION

We consider a MIMO-OFDM broadcast channel with one base station (BS), equipped with $M$ antennas, and $K \geq M$ single-antenna user terminals (UT). MIMO broadcast channels have been widely studied in the recent past (see for example [1], [2], [3], [4], [5]). Under perfect transmitter channel state information (CSIT) at the BS and receiver channel state information (CSIR) at the UTs, its capacity was fully characterized in [5] and efficient resource allocation algorithms have been proposed in order to operate at desired points in the capacity region (e.g., [6], [7], [8]). In the current standardization of the 4-th Generation of wireless communication systems (e.g., IEEE802.16m), MIMO broadcast schemes are going to play a fundamental role in order to achieve high data rates in the downlink. In practice, CSIT must be provided to the BS by some form of feedback.

CSIT feedback schemes are a very active area of research (see for example [9] and the special issue [10] for a fairly complete list of references). In brief, we may identify three broad families: 1) open-loop schemes based on channel reciprocity and uplink training symbols, applicable to Time-Division Duplexing (TDD); 2) closed-loop schemes based on feeding back the unquantized channel coefficients (analog feedback); 3) closed-loop schemes based on explicit quantization of the channel vectors and on feeding back quantization bits, suitably channel-encoded (digital feedback). Closed-loop schemes are suitable for Frequency-Division Duplexing (FDD), where channel reciprocity cannot be exploited. Most if not all present works deal with the case of a frequency-flat channel. In particular, it was recognized that the most important information about the channel vectors consists of their directions. Directional quantization is obtained by using vector quantization codebooks formed by unit vectors distributed on the $M$ dimensional complex sphere. In [11], ergodic achievable rates are analyzed assuming linear zero-forcing beamforming (ZFBF) and random ensembles of spherical quantization codebooks, uniformly distributed on the unit sphere. These results have been extended in [9] to a variety of cases including realistic feedback channels with noise, fading and delay, and to non-perfect CSIR at the UTs obtained by explicit downlink training. In particular, these works show that the sum-rate scales optimally, as $M \log \mathsf{SNR} + O(1)$, provided that the number of quantization bits per UT increases with SNR as $B = \alpha(M-1)\log_2 \mathsf{SNR}$ for some $\alpha \geq 1$. For example, at SNR of $10$ dB a codebook of size 1024 is needed for $M = 4$ antennas, and a codebook of size $2^{24} = 16777216$ is needed for $M = 8$ antennas. Clearly, such channel vector quantizers involve an enormous computational complexity unless some special structure is exploited. Structured spherical vector quantizers for direction quantization have been studied, for example, in [12].

The frequency-selective (OFDM) case is more directly relevant to 4-th Generation wireless systems. A trivial solution consists of operating one independent CSIT feedback per carrier. This solution is suboptimal since it does not take advantage of the fact that the channel vectors at different carriers are correlated. In this paper we compare three channel state feedback schemes for the MIMO-OFDM downlink: analog feedback, digital direction quantized feedback and a new "time-domain" channel quantized feedback inspired by rate-distortion theory. For each scheme we derive the conditions under which the system achieves full multiplexing gain (i.e., the pre-log factor of the sum-rate is equal to $M$). The new rate-distortion inspired scheme takes full advantage of the channel frequency correlation structure and it is shown to outperform the first two. Furthermore, time-domain quantization is by far simpler to implement than complicated spherical vector quantization. In particular, it is seen that no structured codebook design for vector quantization is actually needed for efficient channel state information feedback.

## II. SYSTEM MODEL

For the sake of analytical simplicity, we do not consider users selection based on CSIT feedback information (e.g., as in [13], [14], [15]). Therefore, without loss of generality, we may assume that a set of $M$ out of $K$ users is selected at each time slot according to some scheduling scheme independent of the channel realizations. Also, we assume perfect CSIR at all UTs and focus solely on the CSIT feedback performance. Channels are identically distributed for all users, and spatially independent (no antenna correlation). Therefore, we focus on the description of the scalar channel between any BS antenna and a generic user, dropping antenna and user index for the sake of notation simplicity. A standard assumption in OFDM is that channels behave locally as linear time-invariant finite impulse response filters of length $L$. We assume block-fading channels, constant on blocks of duration $T \gg L$ symbols, and changing according to some ergodic statistics from block to block. In this work we consider zero-delay CSIT feedback and block-by-block estimation. Therefore, we are not concerned with the time-correlation from block to block of the channel (the case of delayed feedback and explicit channel prediction is considered in [9]). Using the standard cyclic-prefix method, blocks of $N = T - L + 1$ information symbols can be transmitted without inter-block interference at the cost of a small dimensionality loss factor of $(1 - \frac{L-1}{T}) \approx 1$, that shall be neglected in the achievable rate expressions of this paper since it applies to all such OFDM schemes in the same way.

After cyclic prefix insertion and removal the resulting channel model is defined by a block transmission of $N$ symbols per transmit antenna, over the $N$ OFDM subcarriers. Letting $\mathbf{h} = [h[0], h[1], ..., h[L-1]]^\mathsf{T}$ denote the discrete-time channel impulse response, the channel in the DFT frequency domain is given by $\mathbf{H} = [H[0], \ldots, H[N-1]]^\mathsf{T}$, where $\mathbf{H} = \sqrt{N}\mathbf{F} \begin{bmatrix} \mathbf{h} \\ \mathbf{0} \end{bmatrix}$ and where $\mathbf{F}$ denotes a unitary $N \times N$ DFT matrix with elements $[\mathbf{F}]_{n,\ell} = \frac{1}{\sqrt{N}}e^{-j2\pi\ell n/N}$, with $n = 0, \ldots, N-1, \ell = 0, \ldots, N-1$. A common assumption consists of modeling the time-domain channel coefficients $h[l]$'s as independent Gaussian random variables $\sim \mathcal{CN}(0, \sigma_l^2)$, where the path variances $\{\sigma_0^2, \ldots, \sigma_{L-1}^2\}$ forms the *Delay Intensity Profile* (DIP) of the channel. We follow this model here, and re-discuss it in Section VI where we show how to take advantage of a more physically motivated channel model. The frequency-domain channel covariance matrix is given by

$$\mathbf{\Sigma}_H = \mathbb{E}[\mathbf{H}\mathbf{H}^\mathsf{H}] = \mathbf{F} \begin{bmatrix} N\mathbf{\Sigma}_h & \mathbf{0} \\ \mathbf{0} & \mathbf{0} \end{bmatrix} \mathbf{F}^\mathsf{H} \tag{1}$$

where $\mathbf{\Sigma}_h = \text{diag}(\sigma_0^2, ..., \sigma_{L-1}^2)$. Furthermore, the diagonal elements of $\mathbf{\Sigma}_H$ are equal to $\sigma_H^2 = \mathbb{E}\left[|H[n]|^2\right] = \sum_{l=0}^{L-1} \sigma_l^2$.

In the MIMO case, the channel from the BS to UT $k$ is defined by the vector discrete-time impulse response $[\mathbf{h}_k[0], \mathbf{h}_k[1], \ldots, \mathbf{h}_k[L-1]]$ where $h_{k,i}[l]$ is the channel coefficient from the BS antenna $i$ to the UT $k$ at discrete-time delay $l$. By applying OFDM modulation and demodulation, the received signal at UT $k$ on the $n$-th subcarrier can be written as

$$y_k[n] = \mathbf{H}_k^\mathsf{H}[n]\mathbf{x}[n] + z_k[n] \tag{2}$$

where $k = 1, ..., K, \ n = 0, ..., N-1, \ \mathbf{x}[n] \in \mathbb{C}^M$ is the transmitted vector of frequency-domain symbols on the $M$ BS antennas, at subcarrier $n$, and $\mathbf{H}_k[n] = [H_{k,1}[n], ..., H_{k,M}[n]]^\mathsf{T}$ is the channel vector of UT $k$ at subcarrier $n$. The average transmit power constraint is given by $\frac{1}{N}\sum_{n=0}^{N-1} \mathbb{E}[|\mathbf{x}[n]|^2] \leq \mathcal{P}$.

For simplicity of analysis, this paper treats only the case of linear Zero-Forcing Beamforming (ZFBF). It is well-known that ZFBF performs at a fixed gap from the optimal capacity achieving strategy under perfect CSIT. Hence, our goal is to find conditions under which ZFBF performs at a fixed rate gap from the perfect CSIT case, which implies fixed rate gap from optimal. For perfect CSIT, the ZFBF transmitted signal at subcarrier $n$ is given by $\mathbf{x}[n] = \mathbf{V}[n]\mathbf{u}[n]$ where $\mathbf{V}[n] \in \mathbb{C}^{M \times K}$ is a zero-forcing beamforming matrix with unit norm columns such that each $k$-th column $\mathbf{v}_k[n]$ is orthogonal to the subspace spanned by $\{\mathbf{H}_j[n] : j \neq k\}$, and $\mathbf{u}[n] \in \mathbb{C}^K$ denotes the vector of coded symbols, independently generated for the different UTs. In high SNR the uniform power allocation yields a fixed rate gap from the optimal (waterfilling) power allocation. Therefore, following [11] and [9], we restrict to this case and let $\mathbb{E}[\mathbf{u}[n]\mathbf{u}[n]^{\mathsf{H}}] = \frac{\mathcal{P}}{M}\mathbf{I}$. Under these assumptions, the achievable rate at each UT $k$ under ZFBF with perfect CSIT is given by

$$R_{k,\mathrm{CSIT}} = \frac{1}{N}\sum_{n=0}^{N-1} \mathbb{E}\left[\log\left(1 + \frac{\left|\mathbf{H}_k^{\mathsf{H}}[n]\mathbf{v}_k[n]\right|^2 \mathcal{P}}{N_0 M}\right)\right] = \exp\left(\frac{N_0 M}{\mathcal{P}\sigma_H^2}\right) E_i\left(1, \frac{N_0 M}{\mathcal{P}\sigma_H^2}\right) \qquad (3)$$

where $E_i(n, x) = \int_1^\infty \frac{e^{-xt}}{t^n}dt, \ x > 0$, is the exponential-integral function.

In the case of non-ideal CSIT, the BS uses the available channel information $\widehat{\mathbf{H}}_k[n]$, $k = 1, \ldots, K$, $n = 0, \ldots, N-1$, and computes the ZFBF matrix $\widehat{\mathbf{V}}[n]$ by treating $\widehat{\mathbf{H}}_k[n]$ as if it was the true channel. The resulting received signal at the $k$-th UT is

$$\begin{aligned}
y_k[n] &= \mathbf{H}_k^{\mathsf{H}}[n]\widehat{\mathbf{v}}_k[n]u_k[n] + \sum_{j \neq k}\mathbf{H}_k^{\mathsf{H}}[n]\widehat{\mathbf{v}}_j[n]u_j[n] + z_k[n] \\
&= a_{k,k}[n]u_k[n] + \sum_{j \neq k}a_{k,j}u_j[n] + z_k[n]
\end{aligned} \qquad (4)$$

where $a_{k,j}[n]$ denotes the coupling coefficient between the user channel $\mathbf{h}_k[n]$ and the beam-forming vector $\widehat{\mathbf{v}}_j[n]$. By following in the footsteps of the achievable rate bound in [9, Theorem 2] we obtain that the achievable ergodic rate for user $k$ is lowerbounded by $R_k \geq R_{k,\mathrm{CSIT}} - \Delta R_k$, where the rate-gap is upperbounded by

$$\Delta R_k \leq \frac{1}{N}\sum_{n=0}^{N-1}\log\left(1 + \frac{\mathbb{E}[|I_k[n]|^2]}{N_0}\right) \qquad (5)$$

with $I_k[n] = \sum_{j \neq k}a_{k,j}u_j[n]$ indicating the multiuser interference term. An upper bound on the rate $R_k$ achievable with Gaussian random coding is also obtained in [9, Theorem 3] by assuming that a genie provides each UT $k$ with exact knowledge of the signal and interference coefficients $a_{k,j}$ for $j = 1, \ldots, M$. This upperbound is referred to as the "genie-aided upperbound" and takes

on the form

$$R_k \leq \frac{1}{N} \sum_{n=0}^{N-1} \mathbb{E} \left[ \log \left( 1 + \frac{|a_{k,k}|^2 \, \mathcal{P}/M}{N_0 + \sum_{j \neq k} |a_{k,j}|^2 \, \mathcal{P}/M} \right) \right] \qquad (6)$$

By dividing both lower and upper bound to the achievable rate by $\log(\mathcal{P}/N_0)$ and letting $\mathcal{P}/N_0 \to \infty$, it is clear that a sufficient condition for achieving full multiplexing gain is that $\Delta R_k$ is a bounded function of the SNR $\mathcal{P}/N_0$.[1] We shall examine this condition under different CSIT feedback schemes in the following sections.

## III. Analog Feedback

Analog feedback consists of sending back the unquantized channel coefficients, transmitted as real and imaginary parts of a complex modulation symbol [16]. We model the feedback channel as AWGN, with the same SNR of the downlink, equal to $\mathcal{P}/N_0$. The more involved case of a fading MIMO multiple-access (uplink) feedback channel is treated, for the frequency-flat case, in [9], [16].

In order to take advantage of the channel frequency correlation, we partition the $N$ subcarriers into $J$ clusters such that $N' = N/J$ is an integer, and feed back only the channel measured at frequencies $n' = iN'$ for $i = 0, 1, ..., J-1$. Each UT transmits its channel coefficients at frequency $n'$ by using $M' \geq M$ feedback channel uses per channel coefficient, for a total of $M'J$ channel uses. This is achieved by modulating the channel vector $\mathbf{H}[n']$ by a $M' \times M$ unitary spreading matrix [9], [16]. After despreading, the noisy analog feedback observation for UT $k$ at frequency $n' = iN'$ is given by

$$\mathbf{g}_k[i] = \sqrt{\beta \mathcal{P}} \mathbf{H}_k[iN'] + \mathbf{w}_k[iN'] \qquad (7)$$

where $\beta = M'/M \geq 1$ and where $\mathbf{w}_k[n'] \in \mathbb{C}^{M \times 1}$ is the AWGN in the feedback channel, with i.i.d. components $\sim \mathcal{CN}(0, N_0)$. The BS performs linear MMSE "interpolation" based on the observation (7) for $i = 0, \ldots, J-1$ and compute the beamforming $\widehat{\mathbf{V}}[n]$ for each subcarrier based on the estimated channel. Since channels are spatially i.i.d., the BS can estimate independently each antenna for each UT. Therefore, without loss of generality, we focus on the side information and estimation of antenna $m$ of UT $k$. By stacking the feedback observations, we form the vector $\mathbf{g}_{k,m} = [g_{k,m}[0], \ldots, g_{k,m}[J-1]]^\mathsf{T}$ given by

$$\mathbf{g}_{k,m} = \sqrt{\beta\mathcal{P}}\mathbf{S}\mathbf{H}_{k,m} + \mathbf{w}_{k,m} \tag{8}$$

where $\mathbf{H}_{k,m} = [H_{k,m}[0], H_{k,m}[1], ..., H_{k,m}[N-1]]^\mathsf{T}$, $\mathbf{w}_{k,m}$ contains the AWGN samples and $\mathbf{S}$ is a $J \times N$ sampling matrix defined by $[\mathbf{S}]_{i,n} = \delta_{n=iN'}$, for $i = 0, \ldots, J-1$ and $n = 0, \ldots, N-1$. By letting $\rho = \beta\mathcal{P}/N_0$, the MMSE estimator of $\mathbf{H}_{k,m}$ from $\mathbf{g}_{k,m}$ is given by

$$\widehat{\mathbf{H}}_{k,m} = \sqrt{\frac{\rho}{N_0}}\boldsymbol{\Sigma}_H\mathbf{S}^\mathsf{H}\left(\rho\mathbf{S}\boldsymbol{\Sigma}_{\mathbf{H}}\mathbf{S}^\mathsf{H} + \mathbf{I}\right)^{-1}\mathbf{g}_{k,m} \tag{9}$$

where $\boldsymbol{\Sigma}_H$ is defined by (1). The corresponding MMSE covariance matrix is given by

$$\boldsymbol{\Sigma}_e = \boldsymbol{\Sigma}_H - \rho\boldsymbol{\Sigma}_H\mathbf{S}^\mathsf{H}\left(\mathbf{I} + \rho\mathbf{S}\boldsymbol{\Sigma}_H\mathbf{S}^\mathsf{H}\right)^{-1}\mathbf{S}\boldsymbol{\Sigma}_H \tag{10}$$

---

[1]This condition is actually stronger, since it requires constant rate gap from optimal. Strictly speaking, full multiplexing gain is achieved if $\Delta R_k$ is $o(\log(\mathcal{P}/N_0))$. However, in the cases analyzed in this work either $\Delta R_k$ is bounded, or it is $O(\log(\mathcal{P}/N_0))$, therefore this option is irrelevant in this context.

Our main result on analog feedback is summarized by the following:

*Theorem 1:* The achievable rate gap of MIMO-OFDM ZFBF with analog CSIT feedback as described above is upperbounded by

$$\overline{\Delta R}_k^{\mathrm{AF}} = \log\left(1 + \frac{M-1}{M}\frac{\mathcal{P}}{N_0}\left[\sum_{i=0}^{L-z-1}\sigma_{[i]}^2 + \sum_{l=L-z}^{L-1}\frac{\sigma_{[l]}^2}{1 + \frac{N\beta\mathcal{P}}{N_0}\lambda_{(l-L+z)}}\right]\right) \tag{11}$$

where $\{\sigma_{[l]}^2 : l = 0, \ldots, L-1\}$ are the DIP components arranged in decreasing order, $z = \min\{J, L\}$ and $\{\lambda_{(i)} : i = 0, \ldots, z\}$ are the non-zero eigenvalues of the matrix $\boldsymbol{\alpha}\boldsymbol{\Sigma}_h\boldsymbol{\alpha}^\mathsf{H}$ arranged in increasing order, where $\boldsymbol{\alpha}$ is the leftmost $J \times L$ block of the matrix $\mathbf{SF}$.

*Proof:* See Appendix I. ∎

In particular, if $z = \min\{J, L\} = L$, as $\mathcal{P}/N_0 \to \infty$ the rate gap is upper bounded by the constant

$$\overline{\Delta R}_k^{\mathrm{AF}} = \log\left(1 + \frac{M-1}{MN}\sum_{l=0}^{L-1}\frac{\sigma_{[l]}^2}{\beta\lambda_{(l)}}\right) \tag{12}$$

A fair comparison of digital and analog CSIT feedback schemes is provided by the achievable rate gap versus the number of CSIT feedback channel uses. For example, the above analog feedback scheme makes use of $M'J$ feedback channel uses. We generally express our results in terms of the normalized number of feedback channel uses per antenna, i.e., through the coefficient $\alpha_{\mathrm{fb}} \geq 1$ such that $\alpha_{\mathrm{fb}}M$ is the total number of feedback channel uses per user per frame.

## IV. Directional Vector Quantization

We consider directional quantization based on random vector quantization (RVQ) codebook ensembles, as in [11]. Each UT has a randomly generated quantization codebook $\mathcal{C} = \{\mathbf{c}_1, ..., \mathbf{c}_{2^B}\}$ consisting of $2^B$ codewords independently and isotropically distributed on the $M$-dimensional unit complex sphere. In order to reduce the number of feedback bits, several current system proposals consider to cluster the subcarriers and feedback the quantized channel only for one representative frequency for each cluster, as done for the analog feedback scheme considered in Section III (see for example [17] in the single-user MIMO-OFDM case). Since it is not clear how to interpolate the direction information over the subcarriers, a common approach consists of *assuming* that the channel is constant over clusters spanning less that the channel coherence bandwidth, and use a piece-wise constant beamforming matrix, computed from the CSIT at the center subcarrier in each cluster. We analyze this "piecewise constant" approach in terms of achievable rate gap. We consider again a grid of $J$ equally spaced frequencies as before. On each such frequency $n'$, the quantization of the channel vector $\mathbf{H}_k[n']$ obeys the rule

$$\widehat{\mathbf{H}}_k[n'] = \arg\max_{\mathbf{c} \in \mathcal{C}} \frac{\left|\mathbf{H}_k^{\mathsf{H}}[n']\mathbf{c}[n']\right|^2}{\left|\mathbf{H}_k[n']\right|^2} \tag{13}$$

The binary indices corresponding the selected quantization codewords $\{\widehat{\mathbf{H}}_k[n'] : n' = iN', i = 0, \ldots, J-1\}$ are fed back to the BS over a perfect (error-free, delay free) digital feedback link, for a total of $B_{\text{tot}}$ feedback bits per UT. The total number of feedback bits per UT per frame is given by $B_{\text{tot}} = BJ$.

Using the MMSE decomposition, the channel vector at a subcarrier $n \neq n'$ in the same cluster of $n'$ can be written as

$$\mathbf{H}_k[n] = \check{\mathbf{H}}_k[n] + \check{\mathbf{e}}_k[n, n'] \tag{14}$$

where $\check{\mathbf{H}}_k[n] = c[n, n']\mathbf{H}_k[n']$ and where we define the channel correlation coefficient between subcarriers $n$ and $n'$ as

$$c[n, n'] = \frac{\mathbb{E}[H_{k,m}[n]H_{k,m}[n']^*]}{\mathbb{E}[|H_{k,m}[n']|^2]} = \frac{\sum_{l=0}^{L-1} \sigma_l^2 \, e^{-j2\pi l(n-n')/N}}{\sigma_H^2}$$

The corresponding MMSE is given by $\sigma_{\check{e}}^2[n, n'] = \sigma_H^2(1 - |c[n, n']|^2)$. The ZFBF matrix $\widehat{\mathbf{V}}[n']$ computed from the quantized channels $\widehat{\mathbf{H}}_1[n'], \ldots, \widehat{\mathbf{H}}_K[n']$ is used for all subcarriers $n$ in the cluster of adjacent frequency indices $\{n'-a, \ldots, n'+b\}$, taken modulo $N$ because of the circulant statistics of the frequency-domain channels, where $a = N'/2 - 1, b = N'/2$ if $N'$ is even and $a = b = \lfloor N'/2 \rfloor$ if $N'$ is odd. Our main result with this form of quantized feedback is given by the following:

*Theorem 2:* The achievable rate gap of MIMO-OFDM ZFBF with digital channel state feedback based on RVQ as described above is upperbounded by

$$\overline{\Delta R_k}^{\text{RVQ}} = \frac{J}{N} \sum_{\delta=-a}^{b} \log\left(1 + \sigma_H^2 \frac{\mathcal{P}}{N_0}\left[|c(\delta)|^2 2^{-\frac{B}{M-1}} + \frac{M-1}{M}(1 - |c(\delta)|^2)\right]\right) \quad (15)$$

where $a, b$ have been defined above and where we define

$$c(\delta) = \frac{\sum_{l=0}^{L-1} \sigma_l^2 \, e^{-j2\pi l\delta/N}}{\sigma_H^2}$$

*Proof:* See Appendix II. ∎

In order to express the total number of feedback bits $B_{\text{tot}}$ in terms of feedback channel uses, we make the optimistic assumption that the feedback link can operate error-free at capacity within the strict one-frame delay constraint. This assumption is justified in light of the achievability results of [9], where it is shown that a rate gap very close to this case can be achieved by using very simple practical codes and taking into account the feedback error probability. It follows that $B_{\text{tot}}$ bits can be transmitted in $\alpha_{\text{fb}}(M-1)$ channel uses, [2] where $\alpha_{\text{fb}} = \frac{B_{\text{tot}}}{(M-1)\log_2(1+\mathcal{P}/N_0)}$. Expressing the rate gap in terms of $\alpha_{\text{fb}}$, we obtain

$$\overline{\Delta R_k}^{\text{RVQ}} = \frac{J}{N} \sum_{\delta=-a}^{b} \log\left(1 + \sigma_H^2 \frac{\mathcal{P}}{N_0}\left[\frac{|c(\delta)|^2}{(1 + \mathcal{P}/N_0)^{\alpha_{\text{fb}}/J}} + \frac{M-1}{M}(1 - |c(\delta)|^2)\right]\right) \quad (16)$$

We observe that the rate gap grows linearly with $\log(\mathcal{P}/N_0)$ unless we let $J = N$. Hence, providing only one direction quantized channel per subcarrier cluster does not take advantage of the channel frequency correlation in an efficient way, since the channel is *not exactly piecewise constant* in frequency. Eventually, for sufficiently large SNR, the channel frequency variations are such that the residual interference will dominate on all frequencies $n \neq n'$. Letting $J = N$ and using $B = B_{\text{tot}}/N$ bits per carrier yields

$$\overline{\Delta R_k}^{\text{RVQ}} \leq \log\left(1 + \sigma_H^2 \left(\frac{\mathcal{P}}{N_0}\right)^{1-\alpha_{\text{fb}}/N}\right) \quad (17)$$

---

[2] For simplicity, we normalize here by $M - 1$ instead of $M$. This is justified by the fact that directional quantization does not include any information on the channel magnitude. Furthermore, our numerical results show that this slight bias against directional quantization does not yield any significant difference in the performance comparisons.

which is bounded (or even vanishing with increasing SNR) as long as $\alpha_{\mathrm{fb}}/N \geq 1$. However, this choice may not be optimal for a given SNR. In practice, for given $\alpha_{\mathrm{fb}}$ and SNR, the system performance can be optimized by choosing the number of clusters $J$. The optimization of $J$ is carried out numerically and generally depends on the operating SNR and on the channel DIP, that determines the correlation coefficient $c(\delta)$.

## V. Time-Domain Quantization

The frequency-domain channel vector $\mathbf{H}_{k,m}$ for a given BS antenna $m$ and UT $k$ can be regarded as a correlated Gaussian source with covariance matrix $\boldsymbol{\Sigma}_H$. Letting $\mathbf{H}_{k,m} = \sqrt{N}\mathbf{F}\mathbf{h}_{k,m}$, where $\mathbf{h}_{k,m}$ is the time-domain channel impulse response for UT $k$ and BS antenna $m$, and noticing that $\mathbf{F}$ is an isometry, it follows that

$$\mathbb{E}\left[\left|\mathbf{H}_{k,m} - \widehat{\mathbf{H}}_{k,m}\right|^2\right] = N\mathbb{E}\left[\left|\mathbf{h}_{k,m} - \widehat{\mathbf{h}}_{k,m}\right|^2\right] \tag{18}$$

where we let $\widehat{\mathbf{H}}_{k,m} = \sqrt{N}\mathbf{F}\widehat{\mathbf{h}}_{k,m}$. It follows that the mean-square distortion for $\mathbf{H}_{k,m}$ is minimized by minimizing the mean-square distortion for $\mathbf{h}_{k,m}$.

### A. Rate-Distortion Limit

Since the components of $\mathbf{h}_{k,m}$ are independent, we are in the presence of a set of $L$ "parallel" Gaussian sources. The rate-distortion function for parallel Gaussian sources and mean-square distortion is given by [18]

$$R(D) = \sum_{l=0}^{L-1} \left[\log_2 \frac{\sigma_l^2}{\gamma}\right]_+ \tag{19}$$

where $\gamma$ is the solution of $\sum_{l=0}^{L-1} \min\{\gamma, \sigma_l^2\} = D$. The number of bits per symbol allocated to the quantization of the $l$-th path is given by $B_l = \left[\log_2 \frac{\sigma_l^2}{\gamma}\right]_+$. Notice that if $\gamma \geq \sigma_l^2$, then $B_l = 0$. This corresponds to the appealing and intuitive fact that more quantization bits should be allocated to the dominant paths. The bit allocation is usually referred to as reverse waterfilling (RWF).

Under the (optimistic) assumption that the CSIT feedback can operate at the rate-distortion limit, our main result with this form of quantized feedback is given by the following:

*Theorem 3:* The achievable rate gap of MIMO-OFDM ZFBF with digital channel state feedback based on time-domain quantization described above is given by

$$\overline{\Delta R}_k^{\text{KL,RWF,Limit}} = \log\left(1 + \frac{M-1}{M}\frac{\mathcal{P}}{N_0}D\right) \tag{20}$$

where $D = \mathbb{E}\left[|\mathbf{h}_{k,1} - \widehat{\mathbf{h}}_{k,1}|^2\right]$ and the number of quantization bits per UT given by $B_{\text{tot}} = MR(D)$ and $R(D)$ is given in (19).

*Proof:* See Appendix III. ∎

The superscript "KL" indicate the fact that the above approach corresponds to quantizing the *Karhunen-Loeve transformed* channel which corresponds to quantizing the time-domain channel vector $\mathbf{h}_{k,m}$, under the assumption of independent coefficients. We wish to study the high-SNR behavior of the rate gap upperbound in Theorem 3, in order to determine conditions under which the full multiplexing gain can be attained. We have the following result:

*Corollary 5.1:* In high SNR regime, the rate gap (20) can be relaxed to:

$$\overline{\Delta R}_k^{\text{KL,RWF,Limit}} = \log\left(1 + \sigma_H^2 \frac{\mathcal{P}}{N_0}\frac{M-1}{M}2^{-R(D)/L}\right) \tag{21}$$

*Proof:* See Appendix IV. ∎

In order to relate the number of feedback bits to the number of feedback channel uses, we make again the assumption that the feedback link can operate error-free at capacity. With a total of $\alpha_{\text{fb}}M = \frac{B_{\text{tot}}}{\log_2(1+\mathcal{P}/N_0)}$ feedback channel uses per UT per frame, we let $R(D) = B_{\text{tot}}/M$ in (21) and obtain

$$\overline{\Delta R}_k^{\text{KL,RWF,Limit}} \leq \log\left(1 + \sigma_H^2\frac{M-1}{M}\left(\frac{\mathcal{P}}{N_0}\right)^{1-\alpha_{\text{fb}}/L}\right) \tag{22}$$

It follows that the rate gap is bounded if $\alpha_{\text{fb}}/L \geq 1$ and it vanishes when the inequality is strict.

### B. Scalar Uniform Quantization

Achieving the rate-distortion limit requires, in general, grouping many source symbols into large blocks and performing optimal vector quantization. On the other hand, the CSIT feedback must have very low delay, and the $L$ channel path coefficients must be quantized and sent back on each slot of $T$ channel uses in order to enable the BS to compute the downlink beamforming matrix. Hence, optimal source coding and low feedback delay are two contrasting issues. Fortunately, it is well-known that simple scalar quantization achieves essentially the same distortion versus SNR behavior, within a constant factor. Here we exploit this fact and consider a simple practical implementation of the above time-domain quantization scheme, where each UT performs uniform scalar quantization on real and imaginary part of its channel coefficients. Real and imaginary parts of each channel coefficient $h_{k,m}[l]$ are quantized independently with $\lfloor B_l/2 \rfloor$ bits, where $B_l$ is obtained, for example, by RWF or by some bit-allocation scheme aimed at minimizing the total distortion. The uniform scalar quantizer $\mathcal{Q}_l$ has $Q_l = 2^{\lfloor B_l/2 \rfloor}$ quantization intervals of size $\Delta_l > 0$ where $Q_l$ is an even integer, with thresholds $0, \pm\Delta_l, \pm2\Delta_l, \dots, \pm(Q_l - 2)\Delta/2$ and midpoint reconstruction levels $\pm\Delta_l/2, \pm3\Delta_l/2, \dots, \pm(Q_l - 1)\Delta_l/2$. The $l$-th path quantizer is obtained by choosing $\Delta_l$ in order to minimize the quadratic distortion

$$\mathcal{D}\left(Q_l, \Delta_l\right) = 2 \sum_{i=0}^{Q_l/2-2} \int_{i\Delta_l}^{(i+1)\Delta_l} \left(\eta - i\Delta_l - \frac{\Delta_l}{2}\right)^2 f(\eta)d\eta + 2 \int_{(Q_l-2)\frac{\Delta_l}{2}}^{\infty} \left(\eta - (Q_l - 1)\frac{\Delta_l}{2}\right)^2 f(\eta)d\eta$$

where $f(\eta) = \frac{1}{\sqrt{\pi\sigma_l^2}}e^{-\frac{\eta^2}{\sigma_l^2}}$. The corresponding rate gap is upperbounded by

$$\overline{\Delta R}_k^{\text{KL,RWF,SUQ}} = \left(1 + \frac{M-1}{M}\frac{\mathcal{P}}{N_0}\sum_{l=0}^{L-1} 2D_l^{\text{SUQ}}\right) \tag{23}$$

where $D_l^{\text{SUQ}} = \min_{\Delta_l > 0} \mathcal{D}\left(Q_l, \Delta_l\right)$. While for any finite $B_l$ the optimization of $\Delta_l$ must be carried out numerically and amounts to a simple line search, we can follow the analysis in [19] in order to capture the high-SNR behavior in closed form. If the total bit budget for quantization is large, we can assume that $B_l \gg 1$ for all $l = 0, \dots, L-1$. Then, our goal is to set $\Delta_l$ such that $\mathcal{D}\left(Q_l, \Delta_l\right) \doteq 2^{-B_l}$, in order to have the same asymptotic behavior of the rate-distortion limit analyzed before. For a real Gaussian source with variance $\sigma_l^2/2$ the following asymptotic upperbound holds [19]

$$\mathcal{D}\left(Q_l, \Delta_l\right) \leq \frac{\Delta_l^2}{12} + (Q_l\Delta_l)^2 P_{\text{over}} + o(\Delta_l^2) \tag{24}$$

where the first term accounts for the so-called "granular distortion" and the second term accounts for the overload distortion, where the overload probability is given by

$$P_{\text{over}} = \int_{(Q_l-2)\frac{\Delta_l}{2}}^{\infty} f(\eta)d\eta \leq \exp\left(\frac{-((Q_l - 2)\Delta_l)^2}{4\sigma_l^2}\right)$$

By choosing $\Delta_l = \sqrt{\frac{4B_l\sigma_l^2}{\log_2 e}}2^{-B_l/2}$ we obtain the desired mean-square distortion behavior that decreases as $D_l^{\text{SUQ}} = \frac{\sigma_l^2}{2}\kappa B_l 2^{-B_l} + o(B_l 2^{-B_l})$ where $\kappa \approx 6$ is a constant independent of $B_l$. In particular, for uniform bit allocation $B_l = B_{tot}/(LM)$ and letting $B_{tot} = \alpha_{fb} M \log_2(1 + \mathcal{P}/N_0)$ we obtain the upperbound

$$
\begin{aligned}
\overline{\Delta R}_k^{\text{KL,RWF,SUQ}} &= \log\left(1 + \kappa\sigma_H^2 \frac{M-1}{M}\frac{\mathcal{P}}{N_0}2^{-B_{tot}/(LM)}\frac{B_{tot}}{LM}\right) \\
&\leq \log\left(1 + \kappa\frac{\alpha_{fb}\sigma_H^2}{L}\frac{M-1}{M}\left(\frac{\mathcal{P}}{N_0}\right)^{1-\alpha_{fb}/L}\log_2\left(1 + \frac{\mathcal{P}}{N_0}\right)\right)
\end{aligned}
\tag{25}
$$

Hence, simple scalar uniform quantization yields a vanishing rate gap as long as $\alpha_{fb}/L > 1$, which coincides with the condition for the rate-distortion limit of Corollary 5.1. On the other hand, this bound is not tight enough to capture the behavior for $\alpha_{fb} = 1$ (indeed, for $\alpha_{fb} = 1$ the bound yields a $\log\log(\mathcal{P}/N_0)$ increase of the rate gap).

In our numerical results we considered the optimization of the bit-allocation $B_l$ subject to the constraint $\sum_{l=0}^{L-1} B_l = B_{tot}$. This is a classical integer programming problem, for which greedy solutions have been considered (e.g., see [20]). We omit the details of the allocation algorithm for the sake of space limitation here. However, it is apparent from the results of Section VII that RWF allocation comes very close to the more computational intensive greedy bit-allocation, and therefore it can be safely used in practice.

## VI. Exploiting the Physical Channel Structure

While most analysis of OFDM systems assumes that the discrete-time channel impulse response $\mathbf{h}$ is formed by $L$ independent Gaussian coefficients, this does not generally hold exactly. The commonly accepted Wide-Sense Stationary Uncorrelated Scattering (WSSUS) fading channel model [21] postulates that multipath components at different delays are uncorrelated. However, the delays of the physical channel are not, in general, integer multiples of the OFDM sampling frequency. In other words, while the continuous-time physical channel may obey the WSSUS model, the corresponding discrete-time channel has correlated coefficients.

In this section we remove this unrealistic assumption and take advantage of the physical channel model. The continuous-time baseband channel impulse response can be written as

$$
c(t; \tau) = \sum_{p=0}^{P-1} c_p(t)\delta(\tau - \tau_p(t))
\tag{26}
$$

where $c_p(t)$ is a stationary Gaussian proper process with first-order distribution $\mathcal{CN}(0, \mu_p^2)$ and $\tau_p(t)$ is the $p$-th path delay [21]. Under the slowly time-varying assumption, $\tau_p(t)$ is assumed to be independent of $t$ for time intervals several order of magnitudes larger than the OFDM symbol duration, while $c_p(t)$ is assumed to be locally time-invariant over the channel coherence time, larger than the OFDM symbol duration.

Let $\psi(t)$ denote the convolution of the transmit and receiving front-end filters (included in the D/A and A/D conversion). Then, the concatenation of filters and physical propagation channel around a reference time $t$ is given by the convolution $h(t; \tau) = \psi(\tau) \otimes c(t; \tau)$. By uniform sampling at rate $1/W$, focusing on an arbitrary reference time $t = 0$ and neglecting the time-dependence because of the locally time-invariance assumption, we arrive at the discrete-time channel impulse response

$$h[l] = \sum_{p=0}^{P-1} c_p \psi\left(\left[l - \tau_p W\right]/W\right) \tag{27}$$

In matrix form, this can be written as $\mathbf{h} = \boldsymbol{\Psi}\mathbf{c}$ where $\boldsymbol{\Psi} \in \mathbb{C}^{L \times P}$, $\mathbf{c} \triangleq (c_0, ..., c_{P-1})^\mathsf{T}$ and $\mathbf{h} \triangleq (h[0], ..., h[L-1])^\mathsf{T}$ as defined before. It is clear that in this case the covariance of $\mathbf{h}$ is not diagonal any longer, and it is given by $\boldsymbol{\Sigma}_h = \boldsymbol{\Psi}\boldsymbol{\Sigma}_c\boldsymbol{\Psi}^\mathsf{H}$ where $\boldsymbol{\Sigma}_c = \text{diag}(\mu_0^2, \ldots, \mu_{P-1}^2)$.

Next, we state our main results on the achievable rate gap for analog feedback and "time-domain" quantized feedback by considering this more realistic channel statistics. We omit the proofs since they follow almost trivially into the footsteps of the previous results. It is however interesting to notice that the main effect of this more refined channel model is to replace $L$ (the length of the discrete-time channel impulse response) by $P$ (the number of physical multipath components). In practice, depending on the shape of $\psi(t)$, we may have $P$ significantly less than $L$. Hence, exploiting the knowledge of the physical channel (in terms of coefficients $\{c_p\}$ and delays $\{\tau_p\}$) yields a clear advantage. In general, we assume that the multipath delays $\{\tau_p\}$ are known to the BS since they vary at a much slower rate and can be reliably tracked by the UTs and fed back at much lower duty cycle. Furthermore, the delays satisfy reciprocity even in FDD, and can be estimated by the BS using the uplink pilot symbols.

For the case of analog feedback, we have:

*Theorem 4:* The achievable rate gap of MIMO-OFDM ZFBF with analog channel state feedback as described in Section III is upperbounded by

$$\overline{\Delta R_k^{\text{AF}}} = \log\left(1 + \frac{M-1}{M}\frac{\mathcal{P}}{N_0}\left[\sum_{p=0}^{P-z-1} \delta_{[p]}^2 + \sum_{p=P-z}^{P-1} \frac{\delta_{[p]}^2}{1 + \frac{N\beta\mathcal{P}}{N_0}\lambda_{(p-P+z)}}\right]\right) \tag{28}$$

where $z = \min\{J, P\}$, $\{\lambda_{(i)} : i = 0, \ldots, z\}$ are the non-zero eigenvalues of $\boldsymbol{\alpha}\boldsymbol{\Psi}\boldsymbol{\Sigma}_c\boldsymbol{\Psi}^{\mathsf{H}}\boldsymbol{\alpha}^{\mathsf{H}}$ arranged in increasing order, and where $\{\delta_{[p]}^2 : p = 0, \ldots, P-1\}$ are the eigenvalues of $\boldsymbol{\Psi}\boldsymbol{\Sigma}_c\boldsymbol{\Psi}^{\mathsf{H}}$ arranged in decreasing order.

In particular, the rate gap is bounded as $\mathcal{P}/N_0 \to \infty$ if $J \geq P$. In this case, it is upper bounded by the constant

$$\overline{\Delta R}_k^{\mathrm{AF}} = \log\left(1 + \frac{M-1}{MN}\sum_{p=0}^{P-1}\frac{\delta_{[p]}^2}{\beta\lambda_{(p)}}\right) \tag{29}$$

The directional RVQ quantization scheme, that operates in the frequency domain under the assumption of piecewise constant channel, cannot take advantage from the physical channel knowledge. As for the "time-domain" approach, since $\mathbf{h}$ is a correlated vector we need to project it onto the appropriate Karhunen-Loeve basis in order to transform it into a set of independent "parallel" Gaussian sources. As a low complexity practical alternative, we consider also the direct quantization of the physical channel path coefficients $\mathbf{c}$.

We decompose $\boldsymbol{\Sigma}_H$ as $\boldsymbol{\Sigma}_H = \mathbf{U}\boldsymbol{\Phi}\mathbf{U}^{\mathsf{H}}$ where $\mathbf{U}$ is a unitary matrix and $\boldsymbol{\Phi}$ is the diagonal matrix of eigenvalues. It follows that $\boldsymbol{\Sigma}_H$ has rank $P < N$, and we let $\phi_0^2, \ldots, \phi_{P-1}^2$ denote its non-zero eigenvalues. Without loss of generality we can take $\mathbf{U}$ to be the tall $N \times P$ matrix of the eigenvectors of $\boldsymbol{\Sigma}_H$ corresponding to the non-zero eigenvalues. First, $\mathbf{H}_{k,m}$ is K-L transformed resulting in $\tilde{\mathbf{c}}_{k,m} = \mathbf{U}^{\mathsf{H}}\mathbf{H}_{k,m}$. Then, RWF bit allocation is applied to the quantization of $\tilde{\mathbf{c}}_{k,m}$. From the application of rate-distortion theory we have:

*Theorem 5:* The achievable rate gap of MIMO-OFDM ZFBF with K-L quantization described above, operating at the rate-distortion limit, is given by

$$\overline{\Delta R}_k^{\mathrm{KL,RWF,Limit}} = \log\left(1 + \frac{M-1}{NM}\frac{\mathcal{P}}{N_0}D\right) \tag{30}$$

where $D = \mathbb{E}\left[|\tilde{\mathbf{c}}_{k,1} - \widehat{\tilde{\mathbf{c}}}_{k,1}|^2\right]$, the number of quantization bits per UT given by $B_{\mathrm{tot}} = MR(D)$, and $R(D) = \sum_{p=0}^{P-1}\left[\log\frac{\phi_p^2}{\gamma}\right]_+$ such that $\gamma$ is the solution of $\sum_{p=0}^{P-1}\min\{\gamma, \phi_p^2\} = D$.

In high-SNR, using an approach similar to what was done in Section V-A and letting $B_{\mathrm{tot}} = \alpha_{\mathrm{fb}}M\log_2(1 + \mathcal{P}/N_0)$, we find the rate gap upper bound in the following simple and appealing form:

$$\overline{\Delta R}_k^{\mathrm{KL,RWF,Limit}} \leq \log\left(1 + \sigma_H^2\frac{M-1}{M}\left(\frac{\mathcal{P}}{N_0}\right)^{1-\alpha_{\mathrm{fb}}/P}\right) \tag{31}$$

As already noticed, this shows that further performance improvement can be obtained by exploiting the structure of the physical channel. In particular, this is the case where $L$ is considerably larger than $P$.

Since the K-L transform requires an SVD of an $N \times N$ matrix, which may be computationally demanding for practical values of the OFDM symbol length $N$, we also consider quantizing directly the time domain coefficients, $\mathbf{c}_{k,m} = [c_{k,m}[0], \ldots, c_{k,m}[P-1]]^{\mathsf{T}}$. Letting $\widehat{\mathbf{c}}_{k,m}$ denote the quantized version of $\mathbf{c}_{k,m}$, we have

$$
\begin{aligned}
\overline{\Delta R}_k^{\mathsf{TQ,RWF,Limit}} &\leq \log\left(1 + \frac{M-1}{M}\frac{\mathcal{P}}{N_0}\mathbb{E}\left[|\boldsymbol{\Psi}\mathbf{c}_{k,1} - \boldsymbol{\Psi}\widehat{\mathbf{c}}_{k,1}|^2\right]\right) \\
&= \log\left(1 + \frac{M-1}{M}\frac{\mathcal{P}}{N_0}D\right)
\end{aligned}
\tag{32}
$$

where $D = \sum_{p=0}^{P-1} \psi_p D_p$ with $D_p = \mathbb{E}\left[|c_{k,1}[p] - \widehat{c}_{k,1}[p]|^2\right]$ and $\psi_p$ is the $p$-th diagonal element of $\boldsymbol{\Psi}^{\mathsf{H}}\boldsymbol{\Psi}$.

The optimal time-domain quantization should consider a modified RWF bit-allocation that minimizes the weighted sum of distortions $D = \sum_{p=0}^{P-1} \psi_p D_p$. This can be straightforwardly done, and also a greedy bit-allocation can be applied to the case of scalar quantization. We omit the details for the sake of space limitation. It is interesting to notice that by applying the geometric-arithmetic mean inequality as in the proof of Corollary 5.1 and noticing that $\sigma_H^2 = \sum_{p=0}^{P-1} \psi_p \mu_p^2$, the rate gap achieved by time-domain quantization is upperbounded by the same expression (31) that holds for the K-L approach. This shows that the use of a K-L transform can only yield marginal improvements to the rate gap for high SNR. Therefore, we conclude that the time-domain quantization of the physical path coefficients provides a very attractive and low complexity solution for the CSIT feedback implementation.

## VII. NUMERICAL RESULTS AND CONCLUSIONS

We considered a MIMO-OFDM system with $M = 4$ transmit antennas at the BS, $K = 4$ single antenna UTs and $N = 64$ carriers. We assumed a discrete-time channel model with 5 independent taps and DIP of $\{0.5, 0.24, 0.17, 0.06, 0.03\}$. Figs. 1 and 2 compare the lowerbounds and upperbounds on the sum rates for different feedback schemes as a function of $\alpha_{\mathsf{fb}}$, that quantifies the amount of total feedback channel uses per frame when SNR= 10dB. The lowerbound on the sum rate is calculated by $R \geq K(R_{k,\mathsf{CSIT}} - \overline{\Delta R}_k)$ where upperbound on the rate gap is computed from (11) for analog feedback, (16) for RVQ, (20) for time-domain quantization and (23) for scalar uniform quantization with both RWF and greedy bit-allocation (GBA). The upperbounds are computed by Monte Carlo simulation. The curve for RVQ corresponds to the optimal value of $J$ obtained numerically for a given $\alpha_{\text{fb}}$.

We notice that RVQ achieves the worst performance. We interpret this fact qualitatively by observing that with RVQ it is not clear how to exploit frequency correlation in an efficient way since the "interpolation" of the direction information over the subcarriers is not easily accomplished. On the other hand, if we augment direction information with (quantized) channel magnitude, we cannot outperform the rate-distortion inspired time-domain quantization, which treats directly the corresponding parallel Gaussian source in terms of mean-square distortion. In terms of order of decay for high SNR, scalar quantization of the time domain channel coefficients yields a very simple scheme that performs very close to perfect CSIT. Furthermore, time-domain scalar quantization is very simple to implement, and requires no complicated construction of spherical codebooks and vector quantization algorithms. Overall, also analog feedback with frequency-domain MMSE interpolation yields very competitive performance at low complexity, although its rate gap remains bounded and does not vanish as SNR increases.

Next we considered the same system with SUI-4 channel model given in [22] and omnidirectional antennas where the continuous-time channel model has 3 taps with path delays $\{0, 1.5, 4\}$ $\mu$s and path variances $\{1, 0.3162, 0.1585\}$. $\psi(t)$ is assumed to be a triangular pulse resulting from convolution of rectangular pulses corresponding to D/A and A/D (sample-hold) with width $1/W = 1\mu$s. The lowerbounds and upperbounds on the sum rate can be computed similar to above. Figs. 3 and 4 compare the lowerbounds and upperbounds on the sum rates for different CSIT feedback schemes as a function of $\alpha_{\text{fb}}$ when SNR= 10dB. We observe that time-domain quantization and K-L domain quantization perform very similar , in accordance with the rate-gap bound analysis done before. This shows that for any practical purpose there is no need of K-L transform.

Finally, we considered the same SUI-4 channel model and compare two cases: 1) the transmit/receive pulse-shaping filter matrix $\mathbf{\Psi}$ is known and 2) the matrix is unknown and the discrete-time channel coefficients are *assumed* to be independent while they are, indeed, correlated. Fig. 5 compares the upperbounds on the sum rates corresponding to different feedback schemes for these two cases as a function of $\alpha_{\text{fb}}$ when SNR= 10dB. As it can be observed, knowledge of masking matrix indeed improves the performance, even for such simple channel model and pulse-shaping filter.

## APPENDIX I

From (1) we have that $\mathbf{S}\boldsymbol{\Sigma}_H\mathbf{S}^{\mathsf{H}} = N\boldsymbol{\alpha}\boldsymbol{\Sigma}_h\boldsymbol{\alpha}^{\mathsf{H}}$ where $\boldsymbol{\alpha}$ is the leftmost $J \times L$ block of the $J \times N$ matrix $\mathbf{SF}$. Using this in (10) we can write

$$
\begin{aligned}
\frac{1}{N}\mathrm{tr}(\boldsymbol{\Sigma}_e) &= \mathrm{tr}\left(\boldsymbol{\Sigma}_h - \rho N \boldsymbol{\Sigma}_h \boldsymbol{\alpha}^{\mathsf{H}}\left(\mathbf{I} + \rho N \boldsymbol{\alpha}\boldsymbol{\Sigma}_h\boldsymbol{\alpha}^{\mathsf{H}}\right)^{-1}\boldsymbol{\alpha}\boldsymbol{\Sigma}_h\right) \\
&= \mathrm{tr}\left(\boldsymbol{\Sigma}_h\left[\mathbf{I} + \rho N \boldsymbol{\Sigma}_h^{1/2}\boldsymbol{\alpha}^{\mathsf{H}}\boldsymbol{\alpha}\boldsymbol{\Sigma}_h^{1/2}\right]^{-1}\right)
\end{aligned}
\tag{33}
$$

where the last line follows from the matrix inversion lemma. Notice that $\boldsymbol{\Sigma}_h$ is diagonal. We let $\{\sigma_{[l]}^2 : l = 0, \ldots, L-1\}$ denote the sorted diagonal elements in decreasing order. Then, we let $\{\lambda_{(i)} : i = 0, \ldots, z-1\}$, with $z = \min\{J, L\}$, denote the non-zero eigenvalues of $\boldsymbol{\alpha}\boldsymbol{\Sigma}_h\boldsymbol{\alpha}^{\mathsf{H}}$ sorted in increasing order. The eigenvalues of the $L \times L$ matrix $\left[\mathbf{I} + \rho N \boldsymbol{\Sigma}_h^{1/2}\boldsymbol{\alpha}^{\mathsf{H}}\boldsymbol{\alpha}\boldsymbol{\Sigma}_h^{1/2}\right]^{-1}$, sorted in decreasing order, are given by

$$
\underbrace{1, \ldots, 1}_{L-z}, \frac{1}{1 + N\rho\lambda_{(0)}}, \ldots, \frac{1}{1 + N\rho\lambda_{(z-1)}}
$$

Now, we use result H.1.g in [23, Ch. 9], stating that for any two $n \times n$ Hermitian positive semidefinite matrices $\mathbf{A}$ and $\mathbf{B}$, we have $\mathrm{tr}\left(\mathbf{AB}\right) \leq \sum_{i=1}^{n}\lambda_i(\mathbf{A})\lambda_i(\mathbf{B})$ where $\lambda_i(\mathbf{A})$ and $\lambda_i(\mathbf{B})$ are eigenvalues of $\mathbf{A}$ and $\mathbf{B}$ sorted in the *same* order. It follows that

$$
\frac{1}{N}\mathrm{tr}(\boldsymbol{\Sigma}_e) \leq \sum_{l=0}^{L-z-1}\sigma_{[l]}^2 + \sum_{l=L-z}^{L-1}\frac{\sigma_{[l]}^2}{1 + N\rho\lambda_{(l-L+z)}}
\tag{34}
$$

For each UT $k$ the channel estimation error on subcarrier $n$ is given by $\mathbf{e}_k[n] = \mathbf{H}_k[n] - \widehat{\mathbf{H}}_k[n]$. Since the noise and the fading process are spatially uncorrelated, we have that $\mathbb{E}\left[\mathbf{e}_k[n]\mathbf{e}_k^{\mathsf{H}}[n]\right] = \sigma_e^2[n]\mathbf{I}$, where $\sigma_e^2[n]$ is the $n$-th diagonal element of $\boldsymbol{\Sigma}_e$ defined in (10). In particular, $\frac{1}{N}\sum_{n=0}^{N-1}\sigma_e^2[n] = \frac{1}{N}\mathrm{tr}(\boldsymbol{\Sigma}_e) = \sigma_e^2$.

We use the rate-gap expression (5), and find

$$
\begin{aligned}
\mathbb{E}\left[|I_k[n]|^2\right] &= \sum_{j\neq k}\mathbb{E}\left[\left|\mathbf{H}_k^{\mathsf{H}}[n]\widehat{\mathbf{v}}_j[n]\right|^2\right]\frac{\mathcal{P}}{M} \\
&= \sum_{j\neq k}\mathbb{E}\left[\left|\widehat{\mathbf{H}}_k^{\mathsf{H}}[n]\widehat{\mathbf{v}}_j[n] + \mathbf{e}_k^{\mathsf{H}}[n]\widehat{\mathbf{v}}_j[n]\right|^2\right]\frac{\mathcal{P}}{M} \\
&= \frac{M-1}{M}\mathcal{P}\sigma_e^2[n]
\end{aligned}
\tag{35}
$$

where the last line follows from the fact that $\widehat{\mathbf{H}}_k^{\mathsf{H}}[n]\widehat{\mathbf{v}}_j[n] = 0$ for any $j \neq k$ from ZFBF, and that $\widehat{\mathbf{v}}_j[n]$ and $\mathbf{e}_k[n]$ are independent, due to the fact that $\widehat{\mathbf{v}}_j[n]$ is a deterministic function of $\widehat{\mathbf{H}}_i[n]$ for $i \neq j$, and $|\widehat{\mathbf{v}}_j[n]|^2 = 1$. Using this in (5) and using Jensen's inequality we obtain

$$
\begin{aligned}
\overline{\Delta R_k^{\mathsf{AF}}} &\leq \frac{1}{N}\sum_{n=0}^{N-1}\log\left(1 + \frac{M-1}{M}\frac{\mathcal{P}}{N_0}\sigma_e^2[n]\right) \\
&\leq \log\left(1 + \frac{M-1}{M}\frac{\mathcal{P}}{N_0}\sigma_e^2\right)
\end{aligned}
\tag{36}
$$

The desired expression (11) follows from (34).

## Appendix II

We compute the variance of the interference term at frequency $n$, where we assume that $n, n'$ are in the same cluster. Using known results on the average distortion of RVQ [11], we can write

$$
\begin{aligned}
\mathbb{E}\left[|I_k[n]|^2\right] &= \sum_{j \neq k}\mathbb{E}\left[\left|\mathbf{H}_k^{\mathsf{H}}[n]\widehat{\mathbf{v}}_j[n']\right|^2\right]\frac{\mathcal{P}}{M} \\
&\overset{(a)}{=} \sum_{j \neq k}\mathbb{E}\left[\left|\left(c[n,n']\mathbf{H}_k[n'] + \breve{\mathbf{e}}_k[n,n']\right)^{\mathsf{H}}\widehat{\mathbf{v}}_j[n']\right|^2\right]\frac{\mathcal{P}}{M} \\
&\overset{(b)}{=} \sum_{j \neq k}\left(|c[n,n']|^2\,\mathbb{E}\left[|\mathbf{H}_k[n']|^2\right]\mathbb{E}\left[\frac{|\mathbf{H}_k^{\mathsf{H}}[n']\widehat{\mathbf{v}}_j[n']|^2}{|\mathbf{H}_k[n']|^2}\right] + \sigma_{\breve{e}}^2[n,n']\right)\frac{\mathcal{P}}{M} \\
&\overset{(c)}{\leq} \sum_{j \neq k}\left(|c[n,n']|^2\,M\sigma_H^2\frac{2^{-B/(M-1)}}{M-1} + \sigma_H^2(1 - |c[n,n']|^2)\right)\frac{\mathcal{P}}{M} \\
&= \sigma_H^2\mathcal{P}\left(|c[n,n']|^2\,2^{-\frac{B}{M-1}} + (1 - |c[n,n']|^2)\frac{M-1}{M}\right)
\end{aligned}
\tag{37}
$$

where (a) follows from (14), (b) follows from the fact $\breve{\mathbf{e}}_k[n,n']$ is zero mean Gaussian independent of $\mathbf{H}_k^{\mathsf{H}}[n']$ and $\widehat{\mathbf{v}}_j[n']$ and that norm and direction of $\mathbf{H}_k[n']$ and iondependent, and (c) from ( Lemma 2 in [11]), the expression of the MMSE in terms of the correlation coefficient $c[n,n']$ and the fact that $\mathbb{E}\left[|\mathbf{H}_k[n']|^2\right] = M\sigma_H^2$ since channels are spatially i.i.d..

The final result follows from (5) and from the fact that $|c(n,n')|^2$ depends only on the difference $\delta = n - n'$ and it is periodic of period $N'$.

APPENDIX III

Let $\mathbf{H}_k[n]$ denote the vector channel of UT $k$ at frequency $n$, and $\widehat{\mathbf{H}}_k[n]$ denote its reconstructed version obtained from the quantization of $\mathbf{h}_{k,1}, \mathbf{h}_{k,2}, \ldots, \mathbf{h}_{k,M}$. By replicating what was done for the analog feedback case, we have that

$$\mathbb{E}\left[|I_k[n]|^2\right] = \frac{(M-1)\mathcal{P}}{M}\sigma_e^2[n] \tag{38}$$

where $\sigma_e^2[n]$ denotes the quantization error per antenna at frequency $n$.

The rate gap for this case can be upperbounded by

$$
\begin{aligned}
\overline{\Delta R_k}^{\text{KL,RWF,Limit}} &\leq \frac{1}{N}\sum_{n=0}^{N-1}\log\left(1 + \frac{M-1}{M}\frac{\mathcal{P}}{N_0}\sigma_e^2[n]\right) \\
&\overset{(a)}{\leq} \log\left(1 + \frac{M-1}{M}\frac{\mathcal{P}}{N_0}\frac{1}{N}\sum_{n=0}^{N-1}\sigma_e^2[n]\right) \\
&= \log\left(1 + \frac{M-1}{M}\frac{\mathcal{P}}{N_0}\frac{1}{N}\mathbb{E}\left[|\mathbf{H}_{k,1} - \widehat{\mathbf{H}}_{k,1}|^2\right]\right) \\
&\overset{(b)}{=} \log\left(1 + \frac{M-1}{M}\frac{\mathcal{P}}{N_0}\mathbb{E}\left[|\mathbf{h}_{k,1} - \widehat{\mathbf{h}}_{k,1}|^2\right]\right) \\
&= \log\left(1 + \frac{M-1}{M}\frac{\mathcal{P}}{N_0}D\right)
\end{aligned} \tag{39}
$$

where (a) follows from Jensen's inequality and (b) from (18).

APPENDIX IV

In high SNR regime we have that a large number of quantization bits per symbol can be used, therefore $\gamma$ becomes small so that, eventually, $\gamma < \min_l \sigma_l^2$ for all $l = 0, \ldots, L-1$. In this limit, all path coefficients are quantized with equal distortion $\gamma$. Therefore, $D = L\gamma$ and from (20) we get

$$\overline{\Delta R_k}^{\text{KL,RWF,Limit}} \leq \log\left(1 + \frac{\mathcal{P}}{N_0}\frac{M-1}{M}L\gamma\right) \tag{40}$$

where $\gamma$ can be obtained from (19) as

$$\gamma = 2^{-R(D)/L}\left(\prod_{l=0}^{L-1}\sigma_l^2\right)^{1/L} \tag{41}$$

Next, we use the geometric-arithmetic mean inequality and write the loser, but more appealing, upper bound

$$\left(\prod_{l=0}^{L-1} \sigma_l^2\right)^{1/L} \leq \frac{1}{L} \sum_{l=0}^{L-1} \sigma_l^2 = \frac{1}{L} \sigma_H^2$$

Using this into (40), we arrive at (21).

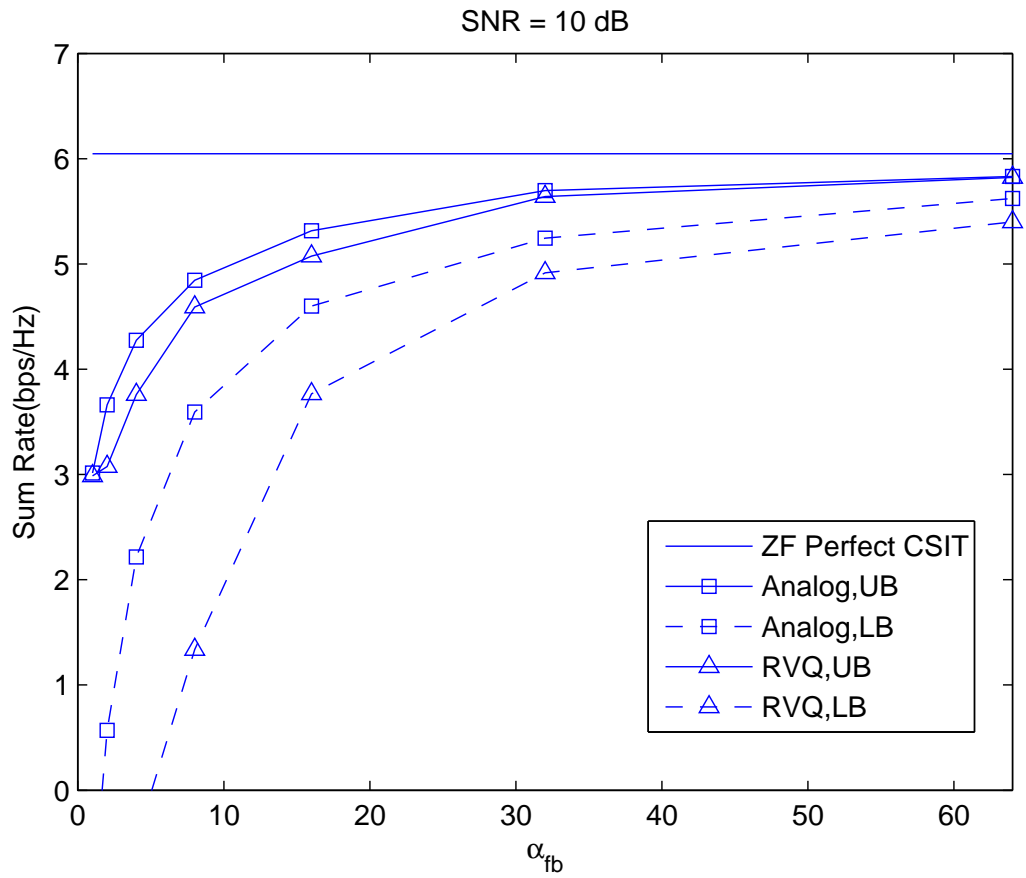

Fig. 1. Comparison of lowerbounds and upperbounds on the sum rate for different feedback schemes with the discrete-time, uncorrelated path channel model when SNR is 10dB.

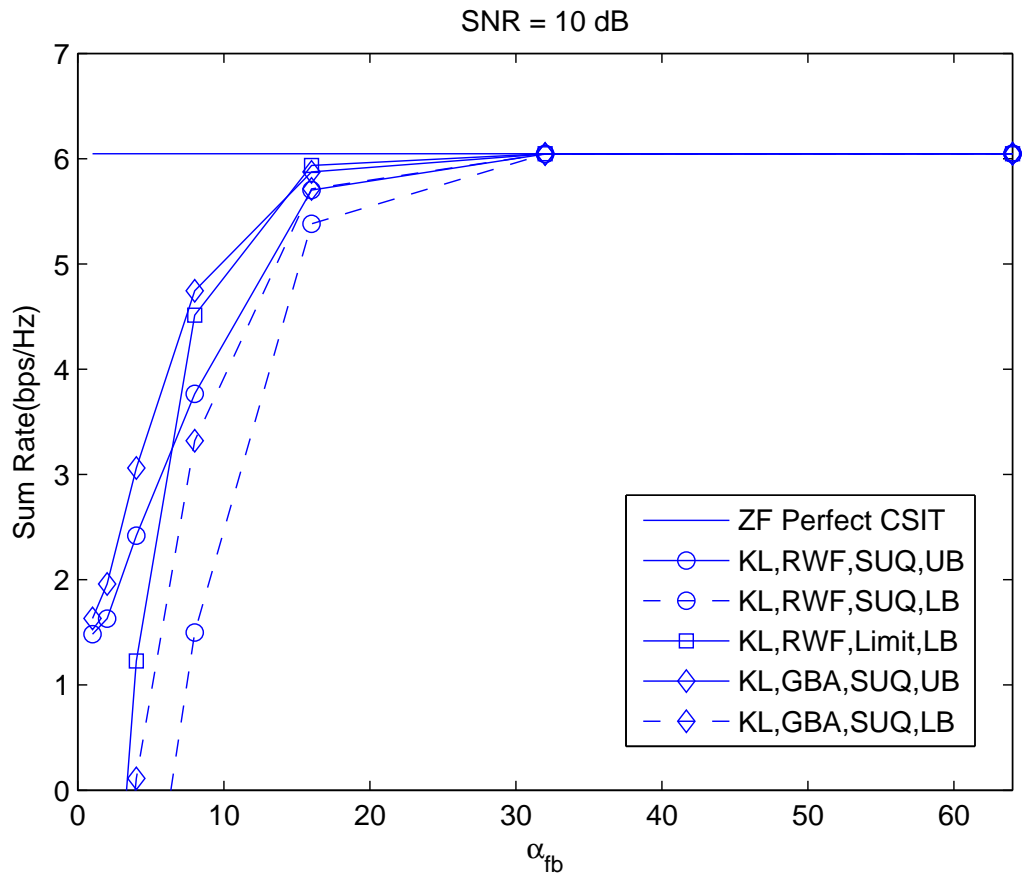

Fig. 2. Comparison of lowerbounds and upperbounds on the sum rate for different feedback schemes with the discrete-time, uncorrelated path channel model when SNR is 10dB.

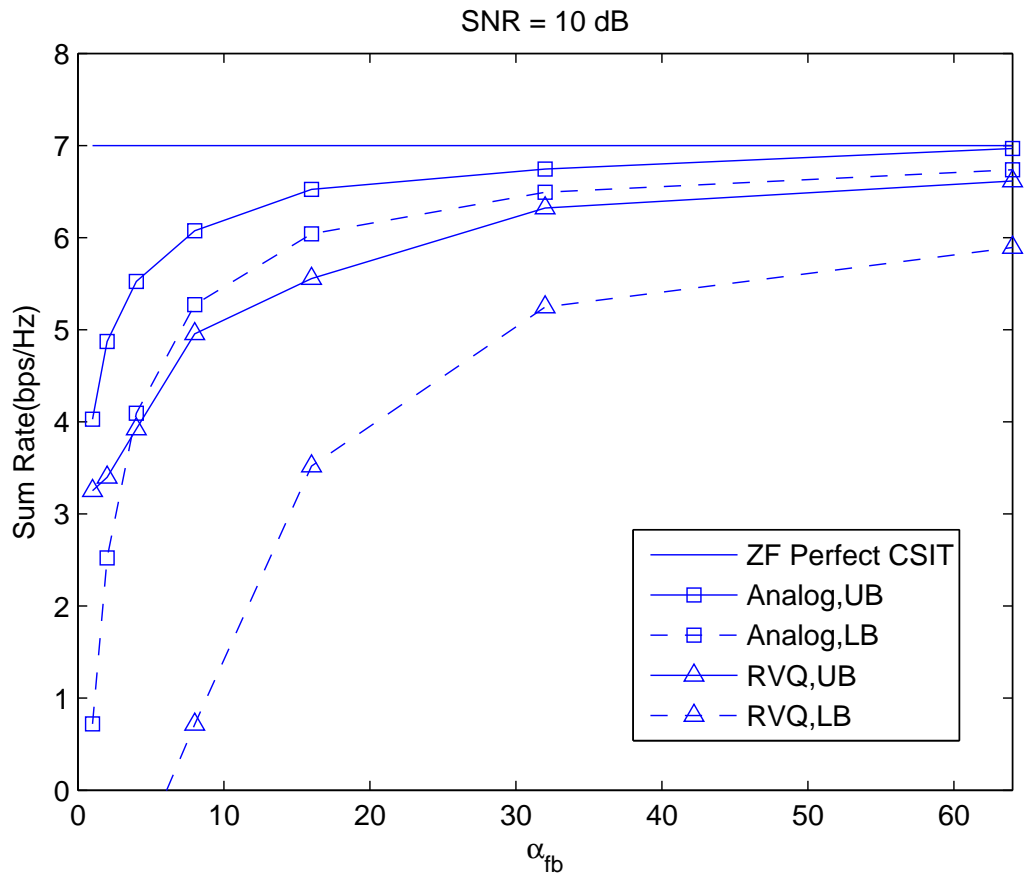

Fig. 3.  Comparison of lowerbounds and upperbounds on the sum rate for different feedback schemes with the continuous-time, uncorrelated path channel model when SNR is 10dB.

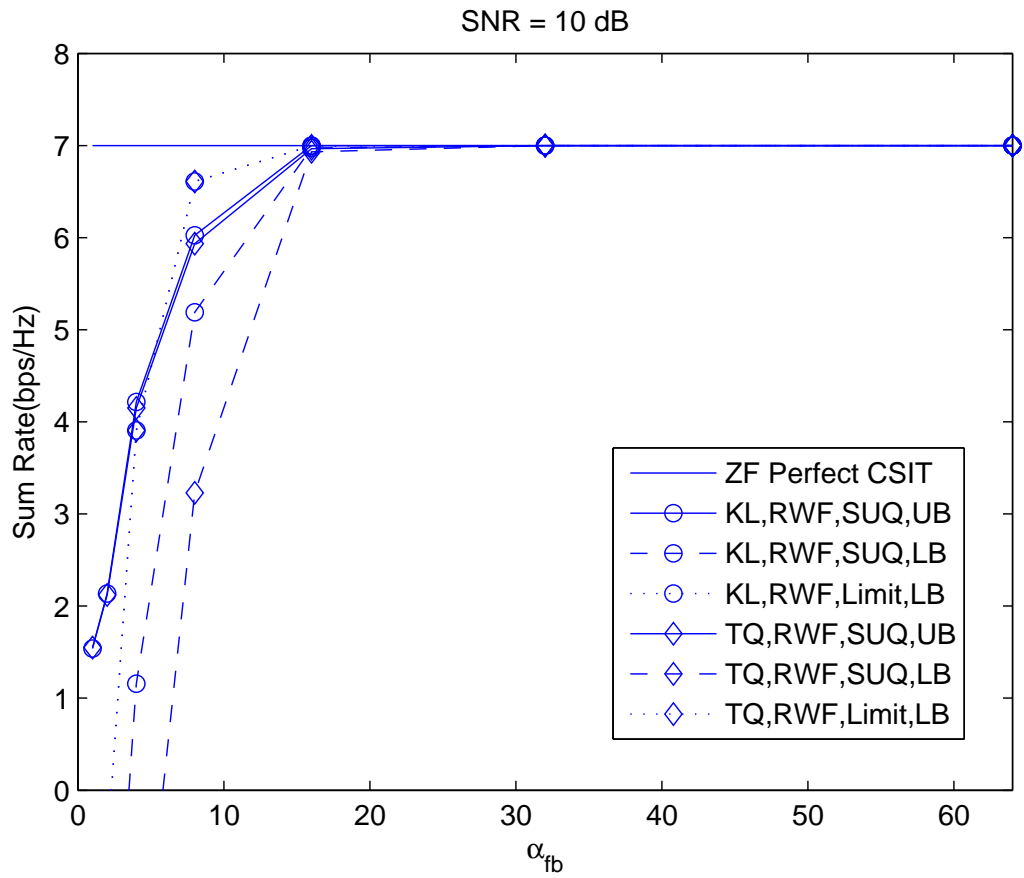

Fig. 4. Comparison of lowerbounds and upperbounds on the sum rate for different feedback schemes with the continuous-time, uncorrelated path channel model when SNR is 10dB.

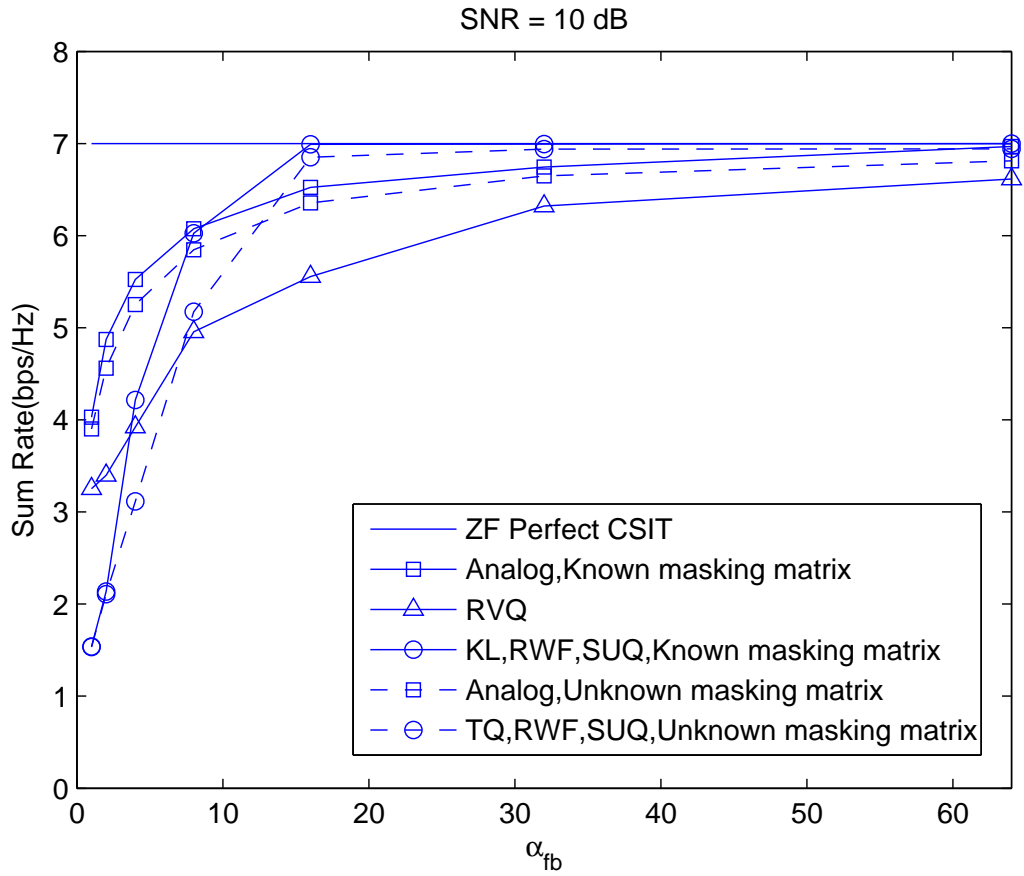

Fig. 5.   Comparison of upperbounds on the sum rate for different feedback schemes with the continuous-time, uncorrelated path channel model for known masking matrix vs. unknown matrix when SNR is 10dB.